 \documentclass[prl,aps,twocolumn,showpacs]{revtex4}
\usepackage[dvips]{graphicx}
\usepackage{dcolumn}
\newcommand{\be}{\begin{equation}}
\newcommand{\ee}{\end{equation}}
\newcommand{\bea}{\begin{eqnarray}}
\newcommand{\eea}{\end{eqnarray}}
\newcommand{\nn}{ \nonumber}

\begin{document}
\topmargin=-10mm

\title{ Local Flattening of the Fermi Surface and Quantum Oscillations in the Magnetoacoustic Response of a Metal}

\author{Natalya A. Zimbovskaya}

\affiliation
{Department of Physics and Electronics, University of Puerto Rico at Humacao, Humacao, PR 00791}

\begin{abstract}
 In the present work we theoretically analyze the effect of the Fermi surface local geometry on quantum oscillations in the velocity of an acoustic wave travelling in metal across a strong magnetic field. We show that local flattenings of the Fermi surface could cause significant amplification of quantum oscillations. This occurs due to enhancement of commensurability oscillations modulating the quantum oscillations in the electron density of states on the Fermi surface. The amplification in the quantum oscillations could be revealed at fitting directions of the magnetic field.
   \end{abstract}

\date{\today}
\maketitle


Oscillations in observables arising due to quantization of conduction electrons motion in strong magnetic fields are well known. They were repeatedly observed in experiments and used as a tool in studies of electron characteristics of metals \cite{one}. At present, these effects are extensively used to study geometrical characteristics of Fermi surfaces in various high-temperature superconductors. Quantum oscillations are specified with contributions from effective cross sections of the Fermi surface (FS) by planes perpendicular to the quantizing magnetic field $ \bf B.$ These are cross sections with minimum and maximum cross sectional areas $ A_{ex}. $ Accordingly, fine geometrical structure of the FS at the effective cross sections and in their vicinities may influence the oscillations. It was already shown that oscillations magnitude could be significantly enhanced when the FS includes nearly cylindrical segments of zero curvature, and some effective cross sections run along these segments \cite{two,three}.
A simple explanation of this effect is that the contributions 
from  nearly cylindrical segments of the FS to 
the electron density of states (DOS) could be sensibly greater 
than those from the rest of the surface. Therefore the 
number of effective electrons associated with a nearly cylindrical cross section is increased which amplifies the oscillations.

It is likely that FSs of some metals include locally flattened segments. Even slight distortion of a Fermi sphere due to the effect of crystalline fields results in the FS flattening at some points \cite{four}. There is an experimental evidence that points of flattening exist on the FSs of cadmium, zinc and potassium \cite{five,six,seven}. When the FS is flattened at some points, this also may bring along an enhancement in the number of effective electrons and amplify the oscillations. To the best of our knowledge, the effect of local flattening of the FS on quantum oscillations was never studied before. We analyze this effect in the present work. As shown below, quantum oscillations could be noticeably strengthened due to the effect of locally flattened segments of the FS, and this could help to locate such segments (if any) in the experiments. The latter brings extra informations concerning fine geometrical features of the FSs. These informations are especially interesting for flattened pieces on the FSs (even small in area) could strongly contribute to the nesting function affecting the strength of the electron-phonon coupling and, consequently, the superconducting temperature in relevant metals. 

To make the presented results more specific  we apply  our analysis to study the effects of the FS local geometry on the quantum oscillations in the velocity of ultrasound waves travelling in a metal. When an ultrasound wave propagates in a metal the crystalline lattice is periodically deformed. It gives rise to electric fields which influence the electrons. Besides, the periodical deformations of the lattice cause changes in the electronic spectrum. Here, we omit these deformation corrections to simplify further calculations.

To proceed we adopt the following energy-momentum relation:
 \be 
E{\bf (p)} = \frac{p_1^2}{2 m_1} \left (\frac{p_y^2 + p_z^2}{p_1^2} \right)^l
+  \left (\frac{p_x}{p_2} \right )^2 \frac{p_2^2}{2m_2}  .
            \ee
This energy-momentum relation could be reasonably justified within a nearly free electron approximation as shown in the recent work \cite{eight}. The corresponding FS is a lens whose radius and thickness are $ p_1 $ and $ p_2 $, respectively. For $ l = 1 $  the energy-momentum relation (1) describes an ellipsoidal FS. In this case $ m_1 $ and $ m_2 $ are the principal values of the effective mass tensor. 

The Gaussian curvature at any point of the lens corresponding to (1) is given by the expression
    \bea 
K{\bf (p)} &=& \frac{l}{m_1 v^4} \left (\frac{p_y^2 + p_z^2}{p_1^2} \right )^{l-1}\nn  \\
  & \times &
\left [(v_y^2 + v_z^2) \frac{\partial v_x}{\partial p_x} +
v_x^2 \frac{l(2l - 1)}{m_1}
\left (\frac{p_y^2 + p_z^2}{p_1^2} \right )^{l-1} \right ].
\nn\\
           \eea

For  $ l > 1,$ the Gaussian curvature $ K(p_x, p_y, p_z)$ turns zero at the points $ (\pm p_2, 0,0)$ coinciding with the vertices of the lens. In view of axial symmetry of the lens, the curvatures of both principal cross-sections turn zero at these points, so the vertices are the points of flattening of the FS. The lens could be a part of a multiply connected FS. For example, electron lenses are included in the FSs of cadmium and zinc, and there are grounds to believe that they are flattened. As usual, we assume the magnetic field to be directed along the $ z $ axis, so, the flattening points at the vertices of the lens belong to an effective cross section.

We consider a longitudinal ultrasound wave travelling along the $y $ axis with frequency  $ \omega $ and wave vector $ {\bf q} = (0,q,0). $ An expression for the wave vector of a sound wave can be written down as follows: 
   \be 
 q = \omega/s + \Delta q. 
   \ee
  Here, $ s $ is the speed of sound in the absence of the magnetic field, and the correction $ \Delta q $ determines the velocity shift $ \Delta s $ and the attenuation rate $ \Gamma:$ 
    \be 
  \frac{\Delta q}{q} = \frac{\Delta s }{s} + \frac{i \Gamma }{2 q}.
   \ee
   This correction is the sum of two terms. The first term $ \Delta q_1 $ describes geometrical oscillations in the ultrasonic attenuation rate and the velocity shift. Such oscillations are very well known in conventional metals (See \cite{one}), as well as in two-dimensional electron systems \cite{nine,ten,eleven}. The effect is controlled with classical magnetotransport mechanisms. Geometric oscillations arise due to the periodic reproduction of commensurability between the diameters of the electron cyclotron orbits and the space period of the external disturbance. In our case the latter is the electric field accompanying the ultrasound wave, while it propagates in a metal. Using the general equations, for the magnetoacoustic response of a metal \cite{twelve}, we can write out the following expression for the contribution from the lens to $ \Delta q_1: $
   \bea 
  \frac{\Delta q_1}{q} &=& - \frac{\omega}{\rho_m s^2} 
\frac{N^2}{g^2} \frac{1}{4 \pi^2 \hbar^3} 
  \nn \\   & \times &
\int d p_z m_\perp (p_z) \sum_k 
\frac{n_{-k} (p_z,-q) n_k (p_z, q)}{\omega + i/\tau - k \Omega (p_z)} .
   \eea
  Here, $\rho_m $ is the density of matter in the lattice; $ N $ is the electron consentration; $ g $ is the electron density of states on the FS in the absence of the external magnetic field; $ m_\perp, \ \Omega $ are the cyclotron mass and the cyclotron frequency for the electrons associated with the lens; $ \tau $ is the scattering time for electrons; and the quantity $ n_k (p_z, q) $ is the Fourier transform in the expansion in terms of an azimuthal angle $ \Phi $ specyfying the electron position at the cyclotron orbit:
  \be 
 n_k (p_z,q ) =  \frac{1}{2 \pi} 
\int_0^{2 \pi} \!  d \Phi \exp \bigg [ i k \Phi
   - \frac{iq}{\Omega} \int_0^\Phi  \!
v_y (p_z, \Phi') d \Phi'\bigg ].
  \ee   
    Another term $ \Delta q_2 $ originates from the quantization of the orbital motion of electrons in strong magnetic fields, and describes quantum oscillations in the velocity shift. Assuming that the cyclotron quantum $ \hbar \Omega $ is small compared to the chemical potential of electrons $ \zeta \ (\gamma^{-1} = (\hbar \Omega/\zeta)^{1/2} <<1), $ we obtain:
   \be 
 \frac{\Delta q_2}{q} = - \frac{1}{2 \rho_m s^2} \frac{N^2}{g} 
n (-q) n(q) \Delta
  \ee
  where $ n(q) = n_0 (0,q), $ and the function $ \Delta $ gives the contribution of the electron lens to the quantum oscillations of the electron DOS:
   \be 
  \Delta = \frac{1}{\gamma} \sum_{r=1}^\infty \frac{(-1)^r}{\sqrt r} \psi_r (\theta) \cos \bigg (\frac{r c A_{ex}}{\hbar |e| B} - \frac{\pi}{4} \bigg) \cos \bigg (\pi r \frac{\Omega_0}{\Omega} \bigg).
   \ee
  Here, $ \psi_r (\theta) = r \theta/ \sinh r \theta ; \ \theta = 2 \pi^2 T/\hbar \Omega_{ex}; \ T $ is the temperature expressed in units of energy, $ \hbar \Omega_0 $ is the spin splitting energy; and $ \Omega_{ex} = \Omega (0). $

At very strong magnetic fields when the characteristic diameter of the cyclotron orbit $ 2 R $ is smaller than the wavelength of the sound $ (qR < 1 )$, we can go to the limit $ q \to 0 $ in the expression (6), and we get $ n_{\pm k} (p_z, \pm q) |_{q=0} = \delta_{k0}. $ Then the semiclassical correction $ \Delta q_1 $ becomes independent of the magnetic field, and magnetic oscillations are fully described with Eq.(7). These are ordinary quantum oscillations originating from the electron DOS oscillations. In moderately strong but still quantizing magnetic field $ (qR \gg 1), $ magnetic oscillations reveal more complex structure.

To show this we evaluate the Fourier transform (6) using the stationary phase method. The main contribution to the integrals over $ \Phi $ in the Eq. (6) comes from the  stationary points at the cyclotron orbit where electrons move in parallel with the electric field created by the acoustic wave. In the chosen geometry these points are the vertices of the lens $ (\pm p_z,0,0) ,$ so we have \cite{thirteen,fourteen}:
     \bea 
 n_{\pm k} (p_z, \pm q) &=& \frac{a(l)}{(qR)^{1/2l}} 
\exp \bigg [\pm i qR \pm \frac{i \pi k}{2} \bigg ]
   \nn \\   &\times &
\cos \bigg [qR - \frac{\pi k}{2} + \frac{\pi}{4 l} \bigg ]
  \eea
  where the cyclotron radius $ R $ depends on $ p_z, $ and the constant $ a (l ) $ is given by:
    \be 
  a_l = \frac{\alpha}{2 \pi l} \Gamma \bigg (\frac{1}{2l} \bigg ) \sqrt{\frac{m_1m_2}{m_\perp^{ex} }}.
   \ee
  Here, the dimensionless factor $ \alpha $ takes on values of the order of unity, $ \Gamma (1/2l) $ is the gamma function.

Substituting the approximation (9) into the expressions (5), and (7) we obtain:
     \be 
 \frac{\Delta q}{q} = \frac{\Delta q_1}{q} + 
\frac{\Delta q_2}{q} = \frac{1}{4 \rho_m s^2} \frac{N^2}{g} 
Y_l (q,\omega ).               
    \ee
  The function $ Y_l (q,\omega )$ introduced in Eq.(11) has the form:
   \bea 
  Y_l (q,\omega ) & = & \frac{a^2(l)}{(qR_{ex})^{1/l}}
 \Bigg  \{ V(\omega) + W (\omega) \cos \bigg[ 2 q R_{ex} + \frac{\pi}{2 l} \bigg ] 
   \nn \\ 
  & + & 2 \cos \bigg[ 2 q R_{ex} + \frac{\pi}{4 l} \bigg ] \Delta \Bigg \}
   \eea
  where
  \bea 
  V (\omega ) & = & \frac{i \pi \omega}{\Omega_{ex}} \coth
 \bigg [\frac{\pi}{\Omega_{ex} \tau} (1 - i \omega \tau) \bigg ], \\  \nn \\ 
  W (\omega ) & = & \frac{i \pi \omega}{\Omega_{ex}} \sinh^{-1}
 \bigg [\frac{\pi}{\Omega_{ex} \tau} (1 - i \omega \tau) \bigg].
     \eea

  There are two oscillating terms included in the expression (12) for $ Y_l (q,\omega). $ The first term originates from the semiclassical dynamical correction $ \Delta q_1$ and depicts commensurability oscillations. The second term gives quantum oscillations superimposed with  the geometrical oscillations. 
The superposition of these two kinds of magnetic oscillations was studied for two-dimensional electron systems \cite{fifteen,sixteen}. Here, we showed that the same effect occurs in  conventional metals.

 \begin{figure}[t]
\begin{center}
\includegraphics[width=8.8cm,height=7.5cm]{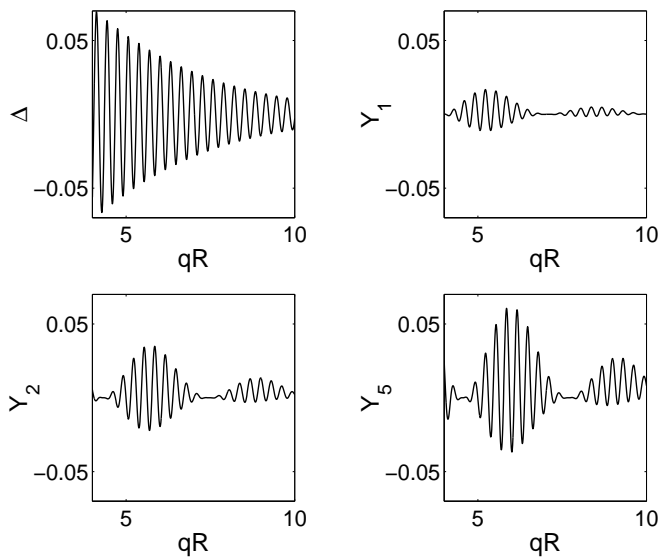}
\caption{ Magnetic oscillations in the electron DOS (top left panel), and the response function $ Y_l (q,\omega) $ associated with the electron lens. The curves are plotted for the ellipsoidal (top right panel) and flattened lens (bottom panels), respectively. In plotting the curves it was assumed that $ \gamma = 10, \ T/\zeta = 2 \times 10^{-3}, \ \omega \tau = 0.1, \ 2 \hbar q p_2 m_{\perp}^{ex} /m_2 A_{ex} = 0.05. $ }
\label{rateI}
\end{center}
\end{figure}

It follows from  (12) that when the FS is flattened in the neighborhoods of the points corresponding to the stationary points at the cyclotron orbit, the magnetic oscillations are noticeably amplified.  In particular, when the ultrasound wave travels across the external magnetic field, the magnitude of quantum oscillations in the velocity shift $ \Delta s /s $ is usually small compared to that of DOS oscillations due to the small gactor $ (qR_{ex})^{-1}, $ as shown in the top panels of the Fig. 1. However, when the FS includes locally flattened segments, the modulated quantum oscillations could reach the same order of magnitude as the DOS oscillations (see the bottom panels of the Fig.1). So the increase in the number of effective electrons originating from the FS local flattening is too small to directly change the DOS oscillations. Nevertheless, it could make an effect on quantum oscillations in the observables by means of amplification of geometrical oscillations modulating the latter.

The amplification of commensurability oscillations due  to  local geometry of the FS can be observed only for a definite choice  of the magnetic field direction with respect to the symmetry axes of the crystal lattice. When the magnetic field is tilted away from such direction the influence of the point of flattening vanishes and the amplitude of the oscillations decreases. Therefore, the amplification of geometric oscillations should exhibit a pronounced dependence on the direction of the magnetic field. Such dependencies were repeatedly discussed for quasi-two-dimentional organic metals \cite{two}.

In summary, in the present work we show  that local flattening of the FS could noticeably influence the magnitude of quantum oscillations in the observables in conventional 3D metals. Unlike the known direct effect of nearly cylindrical segments, the effect of points of flattening on the FS occurs due to amplification of commensurability oscillations modulating DOS quantum oscillations. The effect could be observed in experiments for some particular directions of the magnetic field provided that the external disturbance propagates across the latter. When revealed, this effect could be helpful to discover locations of flattened segments on the FSs. This could be an interesting contribution to the fermiology of high-temperature superconducting materials.

{\it Acknowledgments:}
The  author thanks G. M. Zimbovsky for help with the manuscript. This work was supported in part by NSF Advance program SBE-0123654 and PR Space Grant NGTS/40091.

\end{document}